\documentclass[a4paper,12pt]{article}
\font\header=cmssdc10 at 20pt
\usepackage[ansinew]{inputenc}   
\usepackage[brazil]{babel}
\usepackage{amsfonts}
\usepackage{indentfirst}
  \usepackage[normalem]{ulem}
  \usepackage{graphicx}
  \usepackage{amssymb}
 \usepackage{draftcopy}
  \usepackage{amsmath}
  \usepackage{rotating}
  \usepackage{color}
  \usepackage{setspace}
  \usepackage{fancybox}
 \usepackage{fancyvrb}
	\usepackage{fancyhdr}
	\usepackage[top=0.5cm,left=1cm,right=1cm,bottom=0.5cm]{geometry}
  \usepackage{wrapfig}

	\newcommand{\comenta}[1]{%
	} 

  \usepackage{float}
  \usepackage{type1cm}
\usepackage{eso-pic}
\usepackage{color}
 \usepackage{amsthm,amsmath,amsfonts}
\usepackage{fullpage,enumerate}
\usepackage{tikz}
\usepackage[colorlinks,citecolor=blue,urlcolor=blue]{hyperref}
\usepackage{hypernat}
\usepackage[notref,notcite]{showkeys}
\usepackage{graphicx}

\begin{document}

{\header Testing randomness for cancer risk}

\vskip1cm

Rinaldo B. Schinazi

Department of Mathematics

University of Colorado, Colorado Springs

rschinaz@uccs.edu

\vskip.5cm


\vskip.5cm

Abstract. There are numerous stochastic models for cancer risk for a given tissue. Many
rely on the following two hypotheses. 

1. There is a fixed probability that a given cell division will
eventually lead to a cancerous cell. 

2. Cell divisions are nefarious or not independently of each other.  

We show that recent data on cancer risk and number of stem divisions is consistent
with hypotheses 1 and 2.

\vskip1cm

{\header Theoretical models}

\vskip1cm

Cancer has long been thought as being provoked by successive somatic mutations, see Knudson (2001) for a history of this hypothesis.
A leading hypothesis in the appearance of cancer is that it starts with a single cell division gone wrong.
Usually one assumes that as a result of a mutation during division a precancerous cell appears and then a cascade
of subsequent mutations triggers the appearance of a cancerous cell. There are numerous probability models around
this idea, see for instance Armitage and Doll (1954), Frank et al. (2003), Knudson (2001) and Schinazi (2006). It is believed
that the number of stem cell divisions is particularly relevant to the appearance of cancer, see Cairns (2002).

Whatever the details of the different probability models 
there seems to be two fixed hypotheses. They are:

1. For a given tissue, at each stem cell division there is a fixed probability $p$  that a new cell starts a process
which eventually will yield a cancerous cell. 

2. Cell divisions are nefarious or not independently of each other. 

These hypotheses can be thought as randomness hypotheses for the appearance of cancer.
In this note we are interested in testing hypotheses 1 and 2 and in estimating $p$. 

To be more precise consider a human tissue that undergoes $D$ stem cells divisions over its lifetime. 
Under hypotheses 1 and 2 we have that the risk of cancer $r$ for this tissue is
$$r=P(\mbox{ cancer })=1-(1-p)^D.$$
Depending on the details of the model $p$ may be expressed exactly using mutation probabilities, see Schinazi (2014).

Under hypotheses 1 and 2,  $pD$ is the expected number of times during the lifetime of this tissue that a cell division will eventually lead to a cancerous cell.
That is, $pD$ is the expected count of distinct cancers that affect a given tissue during its lifetime.
It seems reasonable to assume that $pD$ is much smaller than 1. Hence, we have the following approximation
$$r\sim 1-(1-pD)=pD.$$
That is, under hypotheses 1 and 2 one expects a linear relationship between $r$ and $D$. We next test this hypothesis.

\vskip1cm

{\header The test}

\vskip1cm

We use the data provided by Tomasetti and Vogelstein (2015 a), see table S1 in their supplemental material. 
Cancer risk $r$ and total number of cell divisions $D$ are given for 31 cancers. Since we are interested in
testing the randomness of the appearance of cancer we excluded 6 data points that predispose a tissue to cancer.
Namely, we excluded 3 inherited conditions (Colorectal adenocarcinoma with FAP, Colorectal adenocarcinoma with Lynch syndrome and Duodenum adenocarcinoma with FAP),
two tissues infected with a cancer provoking virus (Head and neck squamous cell carcinoma with HPV-16, Hepatocellular carcinoma with HCV) and lung cancer for smokers.
See also the discussion in Tomasetti and Vogelstein (2015 b).
This leaves us with 25 data points. Here are the results:

Pearson's linear correlation between $r$ and $D$ is 0.67.

Linear regression equation is
$$r=1.01\times 10^{-13}D+0.01.$$

We reject the null hypothesis $p=0$ with a P-value $<0.001$.

Hence, an estimate of the parameter $p$ is $\hat p=1.01\times 10^{-13}.$
One expects $p$ to be quite small since it is the probability of eventual appearance
of a cancer cell per division.

\medskip

{\bf Stability of the results}

\medskip

Given that the number of cell divisions $D$ is of order $10^{10}$ and the
cancer risk $r$ is of order $10^{-4}$ one concern is the stability of our estimates.

To check the stability of our results we excluded 5 data points at random 
from the 25 data points and ran the computations. We ran the computations 100 times 
on the remaining randomly selected 20 points. We got an average correlation of 0.69 and an average $\hat p$ of
$1.1\times 10^{-13}.$ These averages are remarkably close to the estimates on the whole sample.
Hence, these results give some confidence in the estimates we obtained.

\vskip1cm

{\header Conclusion}

\vskip1cm

The data is consistent with the idea that cancer
has a large stochastic component. This idea is central to many of the theoretical models for cancer risk.
The correlation between cancer risk and total number of stem cell divisions is surprisingly high.
The estimate for $p$ seems consistent with the hypothesis that $p$ is a product of small probabilities.

\vskip1cm

{\header References}

\vskip1cm

Armitage P. and Doll R. (1954) The age distribution of cancer and a multistage theory of carcinogenesis. British Journal of cancer {\bf 8:} 1-12.

Cairns J. (2002) Somatic stem cells and the kinetics of mutagenesis and carcinogenesis. Proceedings of the National Academy of Sciences {\bf 99:} 10567-10570.
 
Frank S.A., Y. Iwasa and M.A. Nowak (2003) Patterns of Cell Division and risk of cancer. Genetics {\bf 163:} 1527-1532.

Knudson (2001) A.G. Two genetic hits (more or less) to cancer. Nature reviews Cancer {\bf 1: }157-162.

Schinazi R.B. (2006) A stochastic model for cancer risk. Genetics, {\bf 174}, 545-547.

Schinazi R.B. (2014) {\sl Classical and spatial stochastic processes.} (second edition) Birkhauser.

Tomasetti C. and  Vogelstein. B (2015 a) Variation in cancer risk among tissues can be explained by the number of stem cell divisions. Science, {\bf 347}, 78-81.

Tomasetti C. and Vogelstein B. (2015 b) Musings on the theory that variation in cancer risk among tissues can be explained by the number of divisions of normal stem cells.
http://arxiv.org/abs/1501.05035

\end{document}